\documentclass[longauth,letter]{aa} 

\usepackage{graphicx}
\usepackage{xcolor}

\usepackage{txfonts}
\usepackage{comment}
\usepackage{array}
\usepackage{natbib,twoopt}
\usepackage[breaklinks=true]{hyperref} 
\usepackage{xcolor}
\hypersetup{
    colorlinks,
    linkcolor={red!50!black},
    citecolor={blue!50!black},
    urlcolor={blue!80!black}
}
\bibpunct{(}{)}{;}{a}{}{,}             
\makeatletter
  \newcommandtwoopt{\citeads}[3][][]{\href{http://adsabs.harvard.edu/abs/#3}%
    {\def\hyper@linkstart##1##2{}%
     \let\hyper@linkend\@empty\citealp[#1][#2]{#3}}}
  \newcommandtwoopt{\citepads}[3][][]{\href{http://adsabs.harvard.edu/abs/#3}%
    {\def\hyper@linkstart##1##2{}%
     \let\hyper@linkend\@empty\citep[#1][#2]{#3}}}
  \newcommandtwoopt{\citetads}[3][][]{\href{http://adsabs.harvard.edu/abs/#3}%
    {\def\hyper@linkstart##1##2{}%
     \let\hyper@linkend\@empty\citet[#1][#2]{#3}}}
  \newcommandtwoopt{\citeyearads}[3][][]%
    {\href{http://adsabs.harvard.edu/abs/#3}
    {\def\hyper@linkstart##1##2{}%
     \let\hyper@linkend\@empty\citeyear[#1][#2]{#3}}}
\makeatother

\begin{document}

   \title{Astrometric detection of a Neptune-mass candidate planet in the nearest M-dwarf binary system GJ65 with VLTI/GRAVITY}

   \author{
  	    GRAVITY Collaboration\fnmsep\thanks
  	    {
  	    GRAVITY is developed in a collaboration by MPE, LESIA of Paris Observatory / CNRS / Sorbonne Université / Univ.
  	    Paris Diderot and IPAG of Université Grenoble Alpes / CNRS, MPIA, Univ. of Cologne, CENTRA - Centro de
  	    Astrofisica e Gravitação, and ESO. 
  	    }:
  	    R.~Abuter                   \inst{4}            \and
  	    A.~Amorim                   \inst{8,7}          \and
        M.~Benisty                  \inst{3}            \and
  	    J.P.~Berger                 \inst{3}            \and
  	    H.~Bonnet                   \inst{4}            \and
  	    G.~Bourdarot                \inst{1}
       \thanks{
       Corresponding authors:
  	    G.~Bourdarot (\href{mailto:bourdarot@mpe.mpg.de}{bourdarot@mpe.mpg.de}), P.~Kervella (\href{mailto:pierre.kervella@obspm.fr}{pierre.kervella@obspm.fr}) \& O.~Pfuhl (\href{mailto:opfuhl@eso.org}{opfuhl@eso.org}). } \and
        P.~Bourget                  \inst{4}            \and 
  	    W.~Brandner                 \inst{5}            \and
  	    Y.~Clénet                   \inst{2}            \and
  	    R.~Davies                   \inst{1}            \and
        F.~Delplancke-Ströbele      \inst{4}            \and 
        R.~Dembet                   \inst{2}            \and 
  	    A.~Drescher                 \inst{1}            \and
        A.~Eckart                   \inst{6,16}             \and
  	    F.~Eisenhauer               \inst{1,18}         \and
  	    H.~Feuchtgruber             \inst{1}            \and
  	    G.~Finger                   \inst{1}            \and
  	    N.M.~Förster~Schreiber      \inst{1}            \and
  	    P.~Garcia                   \inst{12,7}         \and
        R.Garcia-Lopez              \inst{19,20}        \and
  	    F.~Gao                      \inst{13}         \and
  	    E.~Gendron                  \inst{2}            \and
  	    R.~Genzel                   \inst{1,14}         \and
  	    S.~Gillessen                \inst{1}            \and
  	    M.~Hartl                    \inst{1}            \and
  	    X.~Haubois                  \inst{9}            \and
  	    F.~Haussmann                \inst{1}            \and
  	    T.~Henning                  \inst{5}            \and
  	    S.~Hippler                  \inst{5}            \and
  	    M.~Horrobin                 \inst{6}            \and
  	    L.~Jochum                   \inst{9}            \and
  	    L.~Jocou                    \inst{3}            \and
  	    A.~Kaufer                   \inst{9}            \and
  	    P.~Kervella                 \inst{2}$^{\star\star}$            \and
  	    S.~Lacour                   \inst{2}            \and
  	    V.~Lapeyrère                \inst{2}            \and
  	    J.-B.~Le~Bouquin            \inst{3}            \and
        C.~Ledoux                   \inst{9}            \and
  	    P.~Léna                     \inst{2}            \and
  	    D.~Lutz                     \inst{1}            \and
  	    F.~Mang                     \inst{1}            \and
        A.~Mérand                   \inst{4}            \and
  	    N.~More                     \inst{1}            \and
        M.~Nowak                    \inst{21}           \and
  	    T.~Ott                      \inst{1}            \and
  	    T.~Paumard                  \inst{2}            \and
  	    K.~Perraut                  \inst{3}            \and
  	    G.~Perrin                   \inst{2}            \and
  	    O.~Pfuhl                    \inst{4}$^{\star\star}$          \and
  	    S.~Rabien                   \inst{1}            \and
  	    D.~C.~Ribeiro                \inst{1}            \and
  	    M.~Sadun Bordoni            \inst{1}            \and
  	    J.~Shangguan                \inst{1}            \and
  	    T.~Shimizu                  \inst{1}            \and
  	    J.~Stadler                  \inst{15}         \and
  	    O.~Straub                   \inst{17}         \and
  	    C.~Straubmeier              \inst{6}            \and
  	    E.~Sturm                    \inst{1}            \and
  	    L.J.~Tacconi                \inst{1}            \and
        K.R.W.~Tristram                  \inst{9}            \and
  	    F.~Vincent                  \inst{2}            \and
  	    S.~von~Fellenberg           \inst{16}         \and
  	    F.~Widmann                  \inst{1}            \and
  	    E.~Wieprecht                \inst{1}            \and
  	    J.~Woillez                  \inst{4}            \and
        S.~Yazici                   \inst{1}            \and 
        G.~Zins                     \inst{4}           
    }

    \institute{
  	    Max Planck Institute for Extraterrestrial Physics, Giessenbachstraße 1, 85748 Garching, Germany \and
  	    LESIA, Observatoire de Paris, Université PSL, CNRS, Sorbonne Université, Université de Paris, 5 place Jules Janssen, 92195 Meudon, France \and 	
  	    Univ. Grenoble Alpes, CNRS, IPAG, 38000 Grenoble, France \and
  	    European Southern Observatory, Karl-Schwarzschild-Straße 2, 85748 Garching, Germany \and
  	    Max Planck Institute for Astronomy, Königstuhl 17, 69117 Heidelberg, Germany \and
  	    1st Institute of Physics, University of Cologne, Zülpicher Straße 77, 50937 Cologne, Germany \and
  	    CENTRA - Centro de Astrofísica e Gravitação, IST, Universidade de Lisboa, 1049-001 Lisboa, Portugal \and
  	    Universidade de Lisboa - Faculdade de Ciências, Campo Grande, 1749-016 Lisboa, Portugal \and
  	    European Southern Observatory, Casilla 19001, Santiago 19, Chile \and
  	    Universidade do Porto, Faculdade de Engenharia, Rua Dr. Roberto, Frias, 4200-465 Porto, Portugal \and
  	    Department of Astrophysical \& Planetary Sciences, JILA, Duane Physics Bldg., 2000 Colorado Ave, University of Colorado, Boulder, CO 80309, USA 
  	    \and
  	    Faculdade de Engenharia, Universidade do Porto, rua Dr. Roberto Frias, 4200-465 Porto, Portugal \and
  	    Hamburger Sternwarte, Universität Hamburg, Gojenbergsweg 112, 21029 Hamburg, Germany \and
  	    Departments of Physics \& Astronomy, Le Conte Hall, University of California, Berkeley, CA 94720, USA \and
  	    Max Planck Institute for Astrophysics, Karl-Schwarzschild-Straße 1, 85748 Garching, Germany \and
  	    Max Planck Institute for Radio Astronomy, auf dem Hügel 69, 53121 Bonn, Germany \and
  	    ORIGINS Excellence Cluster, Boltzmannstraße 2, 85748 Garching, Germany \and
  	    Department of Physics, Technical University of Munich, 85748 Garching, Germany \and
        Dublin Institute for Advanced Studies, 31 Fitzwilliam Place, D02,XF86 Dublin, Ireland \and
        School of Physics, University College Dublin, Belfield, Dublin 4, Ireland \and
        Institute of Astronomy, University of Cambridge, Madingley Rd., Cambridge CB3 0HA, UK
    }

   \date{Received 8 February 2024 / Accepted 20 March 2024}

 
  \abstract
  {The detection of low-mass planets orbiting the nearest stars is a central stake of exoplanetary science, as they can be directly characterized much more easily than their distant counterparts. Here, we present the results of our long-term astrometric observations of the nearest binary M-dwarf Gliese 65 AB (GJ65), located at a distance of only 2.67pc. We monitored the relative astrometry of the two components from 2016 to 2023 with the VLTI/GRAVITY interferometric instrument. We derived highly accurate orbital parameters for the stellar system, along with the dynamical masses of the two red dwarfs. The GRAVITY measurements exhibit a mean accuracy per epoch of 50-60 microarcseconds in 1.5h of observing time using the 1.8m Auxiliary Telescopes. The residuals of the two-body orbital fit enable us to search for the presence of companions orbiting one of the two stars (S-type orbit) through the reflex motion they imprint on the differential A-B astrometry. We detected a Neptune-mass candidate companion with an orbital period of $p=156\pm 1\,\mathrm{d}$ and a mass of $m_p=36\pm 7\,M_{\oplus}$. The best-fit orbit is within the dynamical stability region of the stellar pair. It has a low eccentricity, $e=0.1 - 0.3$, and the planetary orbit plane has a moderate-to-high inclination of $i > 30^{\circ}$ with respect to the stellar pair, with further observations required to confirm these values. These observations demonstrate the capability of interferometric astrometry to reach microarcsecond accuracy in the narrow-angle regime for planet detection by reflex motion from the ground. This capability offers new perspectives and potential synergies with Gaia in the pursuit of low-mass exoplanets in the solar neighborhood.}

   
   \titlerunning{Astrometric detection of a planet in GJ65}
   \authorrunning{GRAVITY Collaboration}
   \maketitle
%
\section{Introduction}

\begin{figure*}[ht]
   \centering
   \includegraphics[width=\hsize]{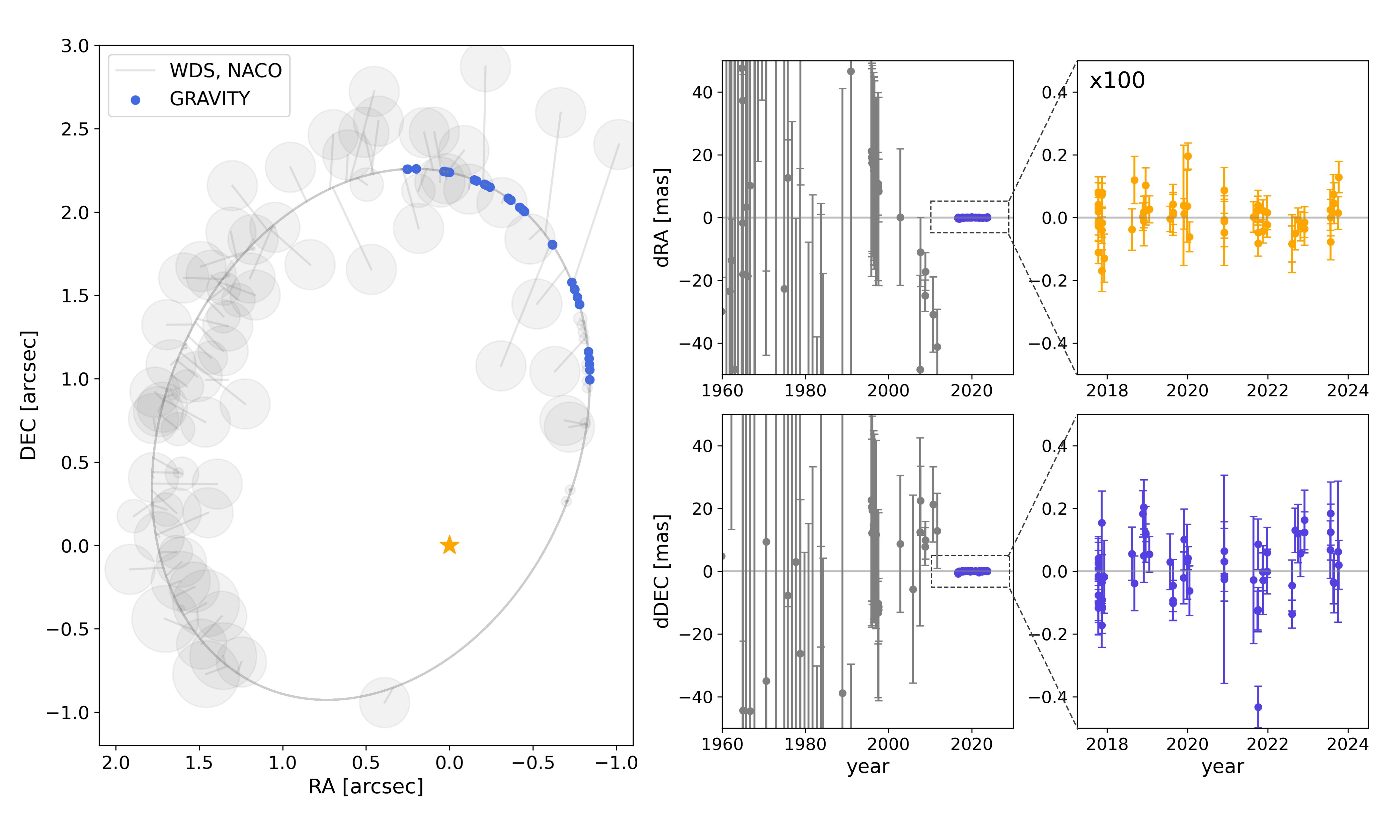}
    \caption{Orbit of GJ65 AB, combining WDS, NACO and GRAVITY data., shown on the left. The residuals of the stellar orbit are on the right.}
         \label{FigStellar}
\end{figure*}

M-dwarfs are favorable targets in the search for signatures of planets thanks to their low stellar masses ($0.1-0.25\,M_{\odot}$), as the presence of planets induces a significantly larger reflex motion on the parent star than for sun-like stars. This makes them ideal targets both for radial velocity and astrometric searches. In our close solar neighborhood, M-dwarfs represent 80\% of the stellar population \citepads{2019AJ....157..216W, 2000ARA&A..38..337C}. 
Thus, the characterization of the properties of the stellar host and their monitoring are of key interest. In that respect, radial velocity and astrometry probe complementary parameter spaces. Radial velocity (RV) measurements are most sensitive to close-in planets, but are ultimately limited to a precision of a few m/s in M-dwarfs due to stellar variability and flaring events \citepads{2021arXiv210406072M}. Conversely, astrometry probes planetary companions at larger separation (typically 0.1 to a few astronomical units, au) and is less sensitive to flare activity, with a typical noise contribution of 10$\mu$au for M-type stars \citepads{2007A&A...476.1389E,2011A&A...528L...9L}. 
In addition, in the case of M stars, the flux contribution of flares decreases by two orders of magnitude in the K band ($2.0\,\mu\mathrm{m}-2.4\,\mu\mathrm{m}$) compared to the UV and optical \citepads{2012ApJ...748...58D}, 
favoring the infrared domain for precision astrometry for these objects. 
The typical astrometric signature of a planet with a few Earth-mass on a 0.1 au orbit around an M dwarf located at 1pc is on the order of 1-10 $\mu$as, below the present limit of space and ground based astrometric instrumentation. 

\begin{table}
\label{table:1}      
\centering                          
\begin{tabular}{p{0.4\hsize}>{\raggedleft\arraybackslash}p{0.4\hsize}}        %
\hline\hline                 
Parameter & Value \\
\hline                        
   Orbital parameters: & \\
   Period $P$ (yr) & $26.38 \pm 0.002$ \\      
   $a$ (AU) & $5.459 \pm 0.002$\\
   $i$ (deg) & $128.0 \pm 0.1$\\
   $e$ & $0.6172 \pm 0.0001$\\
   $\Omega$ (deg) & $325.9 \pm 0.1$\\
   $\omega$ (deg) & $103.2 \pm 0.1$\\
   $T_0$ (mjd) & $41333 \pm 8$ \\
\hline
   Masses: & \\
   $m_A$ ($M_{\odot}$) & $0.122 \pm 0.002$ \\
   $m_B$ ($M_{\odot}$) &  $0.116 \pm 0.002$\\
   $m_A+m_B$ ($M_{\odot}$) & $0.238 \pm 0.003$\\ 
\hline                                   
Parallax$^1$ (mas) & $371.92 \pm 0.42$  \\    
\hline
\end{tabular}
\caption{Orbital parameters of GJ65 A and B obtained with GRAVITY. $^1$ from \citetads{2022A&A...657A...7K}}
\end{table}

In this context, narrow-angle astrometry with interferometry \citepads{2023ARA&A..61..237E} provides a powerful technique for monitoring stellar orbits with very high precision. 
\citetads{1992A&A...262..353S} first discussed the potential of micro-arcsecond precision from the ground with long-baseline interferometry to observe two objects contained in the same isopistonic area. This technique was first demonstrated at PTI \citepads{1999ApJ...510..505C}, achieving ~100$\mu$as accuracy \citepads{2010AJ....140.1657M}, but put on hold after the cancellation of the large space-based NASA mission SIM \citepads{2008PASP..120...38U} and  first-generation ground-based instruments (Keck/ASTRA, \citetads{2010SPIE.7734E..12W}; VLT/PRIMA, \citetads{2006SPIE.6268E..0UD}). The full potential of dual-field interferometry was finally reached with GRAVITY \citepads{2017A&A...602A..94G}, routinely achieving an accuracy of a few $10\,\mu$as on 8m class telescopes for objects as faint as $m_K$=19.5mag \citepads{2018A&A...618L..10G, 2022A&A...657A..82G}.

In the following, we present the astrometric monitoring of GJ65 AB (also known as: Luyten 726-8, UV Ceti+BL Ceti, WDS J01388-1758 AB) we conducted with VLTI/GRAVITY on the 1.8m VLTI Auxiliary Telescopes between 2016 and 2023. GJ65 AB is a binary star located at $d=2.67\,\mathrm{pc}$, and one of the Sun’s nearest neighbors. GJ65 is the class prototype of UV Ceti Type stars \citepads{1949ApJ...109..532L,1949PASP...61..133J}. The separation of the binary ($a=5.4\,\mathrm{au}$) is wide enough to neglect magnetic and tidal interactions between the stars, but its orbital period $P=26.38\,\mathrm{yr}$ is short enough to allow for an extremely accurate determination of the orbital parameters \citepads{2016A&A...593A.127K}. 
This makes GJ65 a prime target in determining the masses of the two stellar components. The stars' properties are very similar to Proxima Centauri \citepads{2016A&A...593A.127K}. The components A and B in GJ65 have nearly identical masses ($\simeq 0.12 M_{\odot}$). No RV detection of a planet has been reported, due to the fast rotation of the two stars and their strong variability through flaring events that limit the accuracy of RV measurements. In astrometry, indications of the presence of a third body from the proper motion anomaly were found from Gaia DR2 and EDR3 but with no clear planet signature \citep{2019A&A...623A..72K, 2022A&A...657A...7K}. Understanding planet formation in close binary systems is still challenging and models show that it is hindered by the dynamical excitation and gravitational instability \citepads{1999AJ....117..621H, 2019Galax...7...84M}. Therefore, obtaining accurate measurements of the stellar orbit and constraining the presence of potential companions is imperative to testing these models.

The outline of the paper is as follows. Section \ref{sec:Observations} describes the observations of our monitoring program and Section \ref{sec:phase-astrometry} presents the principle of dual-field observations used for narrow-angle astrometry with long baseline interferometry, and the performance achieved in the program. In Section \ref{sec:Analysis}, we present the orbital parameters and the analysis of the residuals of the astrometry of GJ65. Finally in Section \ref{sec:Discussion}, we discuss the properties and the dynamical stability of this system, and offer an outlook for future studies.

\section{Observations}
\label{sec:Observations}
We observed GJ65 from 2016 to 2023 with GRAVITY/ATs, with typically 15h of observations per semester. The individual epochs have a total observing time ranging from 1h to 3.3h including overheads. All data were acquired in medium resolution ($R=500$) and SPLIT polarization mode to separate the stellar light in two linear polarizations. Each observation consists of a series of field swaps of about 20 min each, during which the two fibers observe each star successively in order to extract the differential astrometric signal. The principle of measurement in dual-field observations and of swap measurement is detailed in Section \ref{sec:phase-astrometry}. The observation sequence does not require an interferometric transfer function calibrator given that the visibility amplitudes are not used in dual-field astrometry and the phases are self-calibrated by the swapping procedure. 

\section{Phase-referenced astrometry}
\label{sec:phase-astrometry}
Narrow-angle astrometry is enabled by the dual-field design of GRAVITY \citepads{2017A&A...602A..94G}.
In this mode, the interferometric beam combiner is able to simultaneously observe two separated interferometric field of view, with two fibers for each telescope, thereby providing two interference patterns for each baseline. The angular separation of the two objects is measured from the optical path difference (OPD) between the white light fringe of these two fringe patterns, delivering astrometry with a high level of precision that is defined by the interferometric baseline. In order to compensate for additional perturbations on the path between these two channels, a metrology system was implemented \citepads{2012SPIE.8445E..1OG, 2016SPIE.9907E..22L} that measures the OPD from the two-beam combiner to the top of the telescopes. The relevant baseline in astrometric observations is the narrow angle baseline \citepads{2013ApJ...764..109W, 2014A&A...567A..75L}, defined by the end of the metrology point, located on the secondary mirror spider arms of each telescope. The relation between the phase measured by GRAVITY and the angular separation between the targets is then \citepads{2014A&A...567A..75L}:

\begin{equation}
\label{eq:astrom_phase}
\begin{aligned}
\phi &= (\phi_{SC} - \phi_{FT}) - \phi_{\mathrm{MET}}  - \frac{2\pi}{\lambda}D(\lambda_m,\lambda)\, \\
&= \frac{2\pi}{\lambda}(\overrightarrow{s}_B-\overrightarrow{s}_A)\cdot\overrightarrow{B}_{NAB}  + \phi_0(\lambda)
\end{aligned}
\end{equation}

with $\phi_{SC}$ as the phase measured on the science channel of GRAVITY, $\phi_{FT}$ the phase measured on the fringe-tracker, $\phi_{\mathrm{MET}}$ the differential OPD measured by the metrology, $\overrightarrow{B}_{NAB}$ the narrow-angle baseline, $D(\lambda_m,\lambda_{SC})$ the dispersion between the wavelength of the metrology and of the starlight, $\overrightarrow{s}_A$ and $\overrightarrow{s}_B$ the position vector of stars A and B in the plane of sky, $\lambda$ the wavelength of the science channel and $\phi_0(\lambda)$ the zero point of the metrology.

To carry out the phase astrometry, it is necessary to remove $\phi_0(\lambda)$, the constant internal offset of the metrology. In dual-field observations, this can be done either by choosing a reference point to the observations, which will be used as phase reference and subtracted from the astrometry \citepads{2018A&A...618L..10G, 2019A&A...623L..11G} 
or by measuring this zero-point independently by the means of successive swaps between the two fibers. The latter option is used in the case of GJ65. We detail the methodology of the measurement in Appendix \ref{annex1}.

We implemented this method and applied it on the standard reduced files produced by the GRAVITY reduction pipeline \citepads{2014SPIE.9146E..2DL}. The output of the process is finally the relative angular separation $\overrightarrow{s}_B -\overrightarrow{s}_A$ of the A and B components.

\section{Analysis}
\label{sec:Analysis}

\begin{figure}
   \centering
   \includegraphics[width=\hsize]{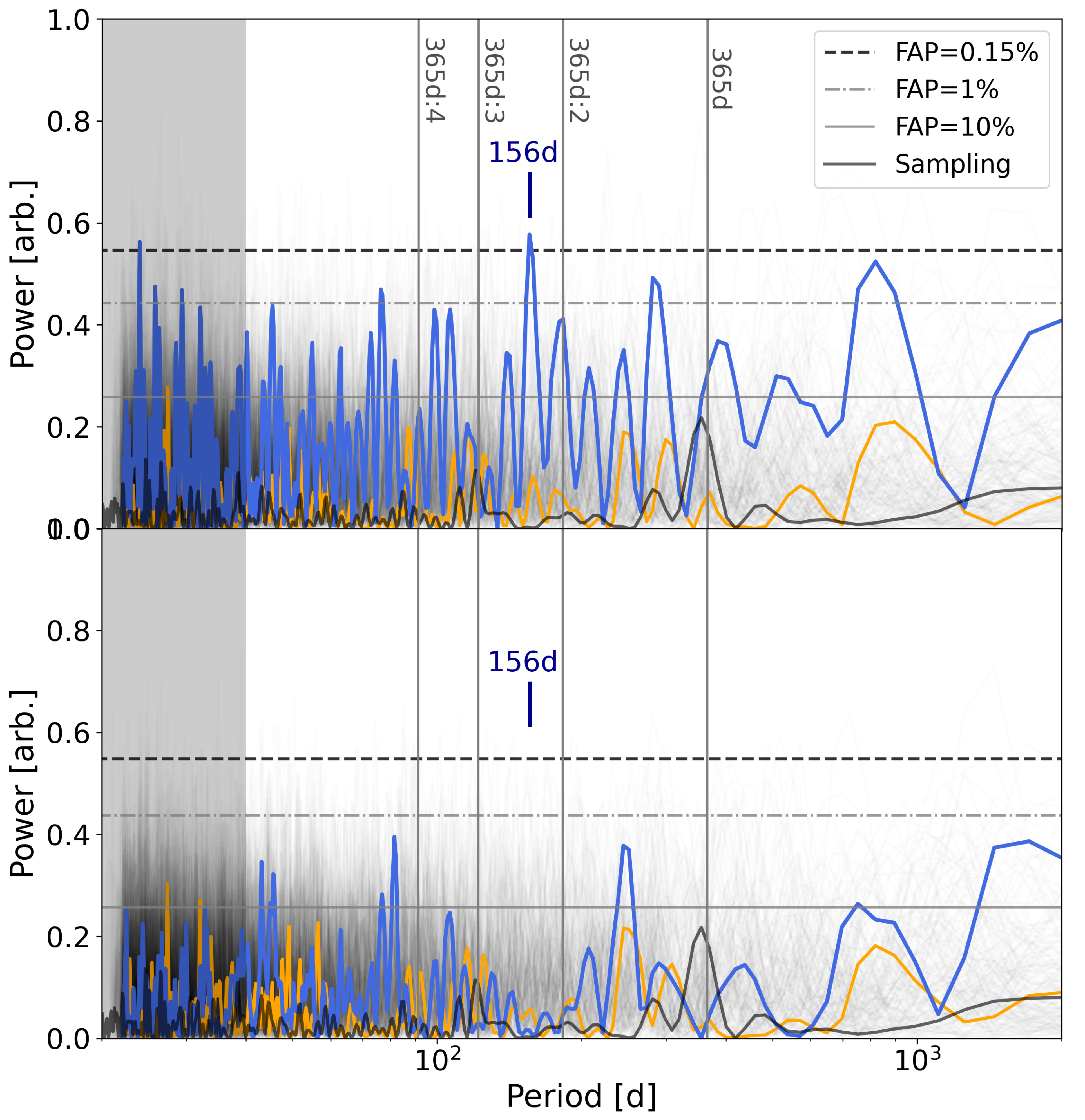}
      \caption{Generalized Lomb-Scargle periodogram. Top: Periodogram in RA (orange) and DEC (blue) computed for the residuals of the binary orbit. Bottom: Residuals after the subtraction of the planet candidate at $p=156\,\mathrm{d}$.
              }
         \label{FigPeriodogram}
   \end{figure}
   
\subsection{Stellar orbit}
\label{sec:stellar_orbit}
We started by fitting the orbital elements of the stellar binary based on the astrometric data.  
The orbital fit includes our new GRAVITY data, complemented by archival astrometry data from WDS catalog and NACO published in \citetads{2016A&A...593A.127K}. The result of the fit is dominated by the GRAVITY data, as shown in Figure \ref{FigStellar} given the much higher accuracy on these points, roughly a factor 100 compared to the archival data.
For each GRAVITY point, the uncertainty and the degree of correlation of the error in RA and DEC are provided as part of the GRAVITY data reduction. The covariance matrix can be reconstructed from these quantities and is integrated in the orbital fit.
The result of the fit is shown in Table \ref{table:1}. We expressed the orbital parameters following the same convention as used in \citetads{2023A&A...674A..34G, 2023A&A...674A...9H}, and shown in Appendix \ref{annex2}. The dynamical mass of the binary is constrained by the combination of the GRAVITY astrometry with UVES radial velocity retrieved from ESO public archive, whose joint fit with the astrometry is shown on Figure \ref{Fig_orbitAnnex_AM} and \ref{Fig_orbitAnnex_RV}.  
This study refines the initial estimate without dual-field astrometry data given in \citetads{2016A&A...593A.127K}, with which they agree. We note the discrepancy in the value of the inclination in \citetads{2016A&A...593A.127K} due to the difference of convention used in our current paper, which is the same as in Gaia DR3.

\begin{figure*}
   \centering
   \includegraphics[width=\hsize]{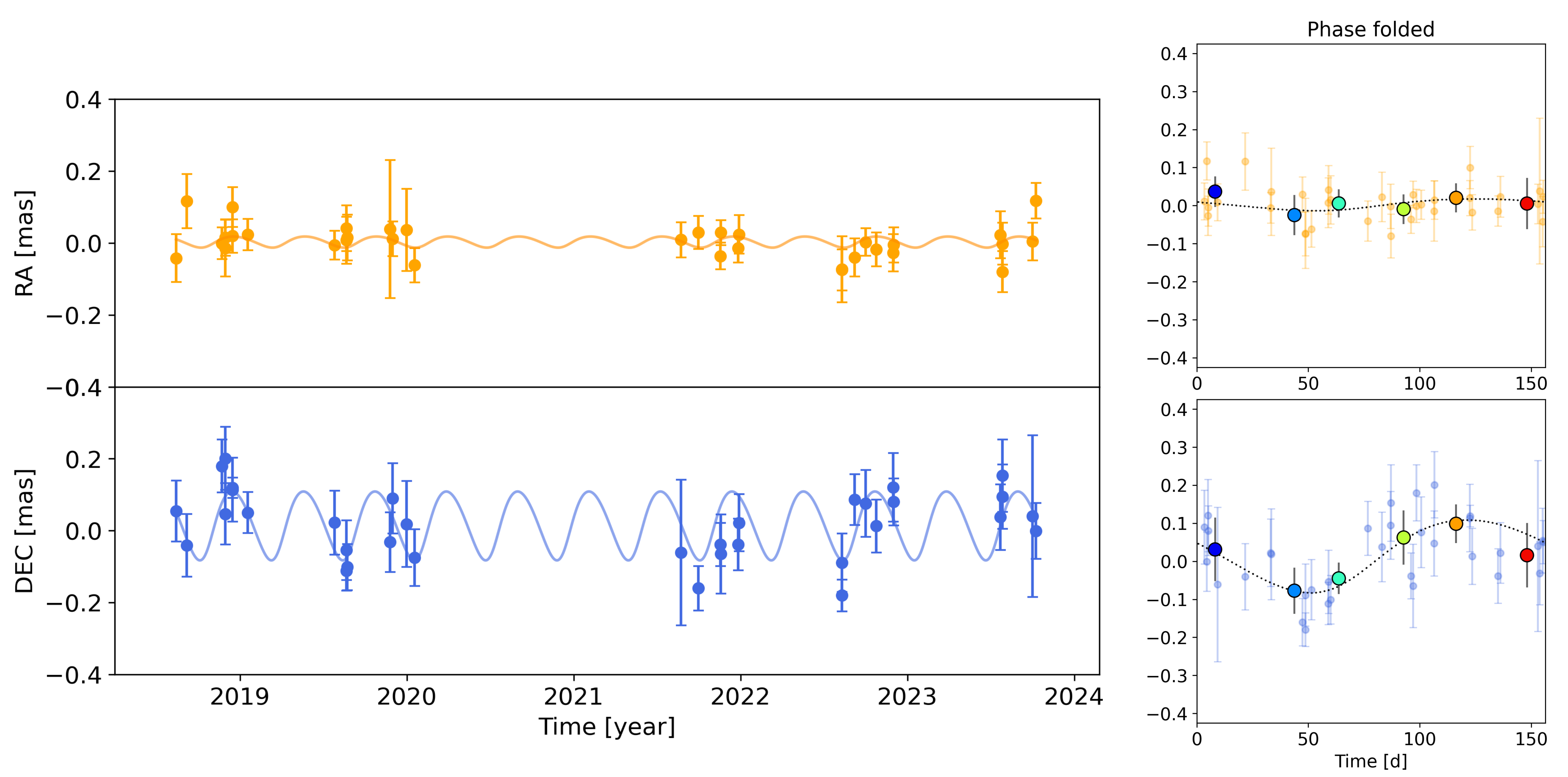}
      \caption{Orbital solution for the best period, $p=156\,\mathrm{d}$ (left). Phase-folded orbit (right) with the average data displayed as coloured dots.
              }
         \label{FigOrbitPlanet}
\end{figure*}

\begin{table}[ht]
\centering                          
\begin{tabular}{p{0.3\hsize}>{\raggedleft}p{0.25\hsize}>{\raggedleft\arraybackslash}p{0.25\hsize}}        %
\hline\hline                 
Parameter & Star B & Star A \\
\hline
Period (day) & $156\pm1$ & $156\pm1$ \\
Mass ($M_{\oplus}$) & $36\pm6$ & $39\pm7$ \\
a (AU) & $0.274 \pm 0.002$ & $0.283 \pm 0.002$ \\
$e$ & $0.27\pm0.21$ & $0.33\pm 0.30$ \\
$i$ (deg) & $89\pm9$ & $88\pm6$ \\
$\Omega$ (deg) & $9^{\star} \pm 7 $ & $8^{\star} \pm 7 $ \\
$\omega$ (deg) & $230\pm36$ & $248\pm30$ \\
\hline
\end{tabular}
\caption{Orbital parameters of the planet for the best period (reduced chi-square $\chi^2_r=1.03$). The host star is indicated for both solutions. \newline $^\star$ Both values ($\Omega,\Omega+180^{\circ}$) are compatible with this parameter given that it is degenerated (see text).} 
\label{table:2}
\end{table}

\subsection{Periodogram}
We computed a generalized Lomb-Scargle periodogram \citepads{2009A&A...496..577Z} on the residuals of the astrometry of the stellar orbit. 
The periodograms of the residuals in right ascension (RA) and declination (DEC) were first computed independently, taking into account their error bars. In addition, we computed the joint periodogram of the signal following the methodology described by \citetads{2006PASP..118.1319C}, which consists of summing the periodograms in RA and DEC, however the detection is mainly driven by the signal in DEC. The result is shown in Figure \ref{FigPeriodogram}.
We used a bootstrapping approach to compute the false alarm probability (FAP) and the detection threshold. To do so, a large number ($N=10^4$) of random periodograms in RA and DEC was generated, then drawn with replacement at the measurement epochs of our sample and following a normal distribution with a zero mean and a standard deviation corresponding to the $1\sigma$ uncertainty of that point. For each draw, a periodogram was generated in RA and DEC. The FAPs were finally estimated by computing the 10\%, 1\% and 0.15\% quantiles of the set of generated periodograms.

Finally, in order to distinguish the impact of the temporal sampling on the final periodogram, we computed the periodogram of the sample replaced by constant values, superimposed on Figure \ref{FigPeriodogram} (gray curve). This "blank" periodogram shows the typical frequencies of our sampling, with a distinct peak at $p=365\,\mathrm{d}$ for the one-year annual periodicity of the program.

The periodogram shows a peak at $p=156\,\mathrm{d}$ above a FAP = 0.15\% threshold. The peak does not overlap with the test periodogram generated on white noise at the temporal sampling of our data set, nor at a harmonic of the one-year period associated to the main period of our program (see Figure \ref{FigPeriodogram}). For the moment, we have discarded the peaks with a period <30d, for which we consider our temporal sampling is not sufficient.

\subsection{Orbit fitting}
In a second step, we performed an orbital fit of the residuals to obtain a second detection of candidate, using a methodology that is different from the Lomb-Scargle periodogram, and to determine the orbital parameters of the companion planet.
Our astrometric model describes the reflex motion of a companion in Keplerian motion around one of the components of the system. With respect to both our stellar orbit and the planetary orbit, the four orbital elements ($a$, $i$, $\Omega$, $\omega$) can be replaced by a linear model using the Thiele-Innes elements.

Firstly, we performed an orbital fit over a grid of fixed periods and fit all the other orbital parameters. In this way, we fixed the period, which is the most non-linear parameters of the fit. Since P is fixed, the fit is left with only two non-linear parameter ($e$, $t_0$) to solve the Kepler equation, and the four Thiele-Innes parameters (A, F, B, G), which are linear coefficients of the elliptical rectangular coordinates ($X,Y$). This method allows us to make the fit robust by limiting the number of non-linear parameters. The covariance matrix of the GRAVITY data, assuming the noise is random Gaussian, is taken into account in the fit as:

\begin{equation}
    \chi^2 = \left(\overrightarrow{y}-\overrightarrow{y}_{\mathrm{model}}\right)^T\Sigma^{-1}\left(\overrightarrow{y}-\overrightarrow{y}_{\mathrm{model}}\right)
\end{equation}

Finally, the Campbell parameters were deduced from the Thiele-Innes parameters, for which we again used the same convention as in \citetads{2023A&A...674A..34G, 2023A&A...674A...9H}.
The result of the grid search over periods converges to the same period $p=156\,\mathrm{d}$ as the Lomb-Scargle periodogram, as shown in Fig. \ref{FigPeriodSearch}. 
The final orbital solution is obtained by allowing all parameters to vary, and converges to a Neptune-mass planet with mass $35-40\,M_{\oplus}$. The result for the best period and the phase folded orbit are shown on Fig \ref{FigOrbitPlanet}. The orbital parameters of the planet are shown in Table \ref{table:2}. 

We note that the fit of the orbit is degenerate to four solutions in total: two possible host stars, and two possible longitudes of the ascending nodes ($\Omega,\Omega+180^{\circ} $) of the planet. However, the prograde solutions (two out of the four in total) are favored from a dynamical point of view, and are the ones shown in Table \ref{table:2}. The degeneracy on the host star could be lifted by measuring the absolute astrometry of the binary, and the degeneracy of the longitude of the ascending nodes by measuring the radial velocity of the host star caused by the motion of the planet.
However, we note that the location of the planet in the plane of sky is known within two possible solutions (degeneracy of the host star), the degeneracy of the ascending node does not affect the knowledge of the location of the planet in the plane of sky (only $z$). The projection of the GJ65 system onto the sky is shown in Fig \ref{Fig_orbitAnnex_Planet}.

\subsection{Additional companions and complementary RV data}
Our orbital solution would lead to a typical RV amplitude of $\approx 20\,\mathrm{m/s}$. We used UVES data to constrain the stellar orbit (see Sec \ref{sec:stellar_orbit}) with a typical accuracy of the order of $\pm$100 m/s; however this is not sufficient to constrain a planetary companion. 
The two components of GJ65 are difficult targets for high precision RV measurements. 
Firstly, their very strong magnetic activity and fast rotation results in spotted and time-variable apparent surfaces \citepads{2017MNRAS.471..811B} that affect the spectral line profiles. Secondly, their relatively small angular separation (Fig. \ref{FigStellar}) induces crosstalk between the two stars for ground-based, seeing-limited spectrographs. 

We note the presence of peaks at shorter periods $\approx 20 \, \mathrm{d}$ both in the Lomb-Scargle periodograms and the period search. Such an astrometric signal would correspond to a Saturn-mass planet on a < 0.1 AU orbit. Here, we limited our analysis to the periods > 30 d due to the temporal sampling of our monitoring under one month. However, the typical RV amplitude of this signal is $\approx 100\, \mathrm{m/s}$, which could be confirmed by RV data. As a follow-up to the <30 d and the 156d candidates, additional RV observations would provide complementary information in this system. The NIRPS high-stability spectrograph at La Silla \citepads{2017Msngr.169...21B} could be a prime instrument for GJ65 AB, thanks to its adaptive optics system which allows for the flux to be isolated from each components of the binary, and in the longer term the high-precision ELT/ANDES instrument.

\section{Discussion}
\label{sec:Discussion}
The candidate exoplanet that we detect in GJ65 is the second nearest after Proxima Centauri’s exoplanet system at the date of writing\footnote{\url{exoplanetarchive.ipac.caltech.edu}}. It thus offers unique opportunities for the further characterization of this exoplanet.
The astrometric detection allows for the mass, inclination and orbital parameters of the candidate to be derived. The location of the planet can be compared with the expected orbital stability region of test particles as derived by \citetads{1999AJ....117..621H} (see Eq. 1 of that paper; Appendix \ref{sec:stability_annex}). The orbital parameters of GJ65AB in Table \ref{table:1} allow us to derive a critical semi-major axis of $a_c= 0.46\, \mathrm{au}$. The exoplanet candidate’s orbit lies within this region, with a semi-major axis $a\simeq 0.283 \,\mathrm{au}$ for the largest value (Table \ref{table:2}). This candidate could thus have formed and evolved within the region of stability. The formation of planets at relatively short separation binaries ($<10\,\mathrm{au}$) is challenging due the dynamical interactions in these systems and the tidal truncation of the protoplanetary disks in early phases of formation, as a disk truncation below the snow line prevents the formation of giant planets \citepads{2005MNRAS.359..521P}. In GJ65, the snowline is located well within the stability region of the binary $\sim 0.04\,\mathrm{au}$ \citepads{2005ApJ...626.1045I}. This is a priori compatible with the formation of a Neptune type planet in this region (Table \ref{table:2}). The current detection thus provides an important example to study the formation mechanism of these systems, although a complete dynamical and evolutionary study of GJ65 is out of the scope of this Letter. Interestingly, GJ65AB shares strong similarities with Alpha Centauri AB from the standpoint of stability, since these two systems share almost identical mass ratio and eccentricity, which are, at first order, the sole parameters  with the semi-major axis entering in $a_c$. GJ65 ($a=5.45\,\mathrm{au}$) is scaled down by a factor 4.2 with respect to Alpha Cen AB ($a=23.52\,\mathrm{au}$), which could form a basis for theoretical long-term stability studies of these two objects \citepads{2016AJ....151..111Q, 2024arXiv240116003C}.

The inclination of the planet's orbit can be compared to the inclination of the stars' rotation axes and that of the binary star orbit. The inclinations of the rotation axes of GJ65A and GJ65B were measured through spectro-polarimetric observations by \citetads{2017MNRAS.471..811B}, who derived inclinations of $i_A = 60\pm 6 ^{\circ}$ and $i_B = 64\pm 7 ^{\circ}$ ($0<i<90 ^{\circ}$), respectively, modulo $180^{\circ}$. They can therefore be either $i=60^{\circ}$ or $i=180-60=120^{\circ}$. The first scenario ($i=60^{\circ}$) would indicate a $\approx 180^\circ$ misalignment between the spin and the orbital plane of the binary, that is, a retrograde orbital configuration. This appears particularly unlikely in GJ65AB, considering that both stars are fast rotators with consistent rotation axis inclinations.
We therefore favor an inclination of $120^{\circ}$ for the stellar rotation axes, as it corresponds to a spin-orbit alignment of star and the binary.
In this configuration, the inclination of the planet's orbital plane ($i_P \approx 90^\circ$; Table~\ref{table:2}) with respect to the binary star's and the two stars' equatorial plane is $\Delta i \approx 30 \pm 10^\circ$.
A similarly high relative inclination is known in several multi-planet systems between planet orbits as, for instance, $\pi$\,Men \citepads{2020A&A...640A..73D,2022AJ....163..223H}, $\nu$\,And \citepads{2010ApJ...715.1203M} or Kepler-108 \citepads{2017AJ....153...45M}.
While the configuration is different in GJ65 with two stars A and B instead of multiple planets, the high eccentricity of their orbit ($e \approx 0.62$; Table~\ref{table:1}) could induce the Kozai-Lidov effect, thereby raising the planet's inclination \citepads{2016ARA&A..54..441N,1962AJ.....67..591K,1962P&SS....9..719L}.
Testing this hypothesis would benefit from follow-up observations to improve the precision of the inclination of the planetary orbit, as well as complementary measurements of the position angle and inclination of the stars' rotation axes, for example, using spectro-interferometry \citepads{2009A&A...498L..41L}.

The immediate vicinity of GJ65 raises the question of characterizing this system with direct imaging. The age of the system was estimated to approximately $\sim 1\,\mathrm{Gyr}$ from mass-luminosity isochrones by \citetads{2016A&A...593A.127K}, based on the evolutionary models for low-mass stars in \citetads{2015A&A...577A..42B}. Given this system is relatively old, the intrisic emission of the planet will be extremely low. The luminosity of the planet will be given by the reflected light contrast $\mathrm{c}=g(\phi)\cdot \,A \cdot \left(\frac{R_p}{a}\right)^2$ with $A$ as the albedo, $g(\phi)$ the phase function on the orbit, and $R_p$ the planet radius. The estimated contrast is $\mathrm{c}=10^{-6.3}$  at a projected separation of 90mas, assuming $g(\phi).\,A=0.4$, and $R_p=7\,R_{\oplus}$ based on the mass-radius relation of \citetads{2017ApJ...834...17C}. This value of contrast is at the limit of current observation capabilities, but could be a prime target for direct high-contrast observations with VLTI/GRAVITY+ and its planned extreme adaptive optics \citepads{2022Msngr.189...17A}, as well as the upcoming ELT instruments \citepads{2018SPIE10702E..1SD, 2022SPIE12185E..5LC, 2023arXiv231117075P}. 
In addition, the knowledge of the orbital parameters of the planet would allow for an astrometry-informed pointing for GRAVITY+ and direct imaging observations from ground and space. 
Gaia DR4 will be useful to further constrain this system and potentially provide absolute astrometry to lift the degeneracy which of the two stars host the candidate planet. However, the astrometric performance may be more affected by flares in the visible compared to infrared observations. The monitoring of GJ65 with follow-up GRAVITY observations will be important in further constraining the orbital parameters of the candidate planet and the possible presence of additional planets. These observations
demonstrate the ability to use astrometry to detect planets around low-mass stars, and could be applied to the monitoring of nearby $<25$pc system to constrain the presence of giant planets down to Neptune mass. In the longer term, the extension of ground-based interferometers to micro-arcsecond precision could enable to probe Earth-mass planets in our close solar neighborhood.

\begin{acknowledgements}
We thank the anonymous referee for his comments which improved the content of this Letter.
We are very grateful to our funding agencies (MPG, ERC, CNRS [PNCG, PNGRAM], DFG, BMBF, Paris Observatory [CS, PhyFOG], Observatoire des Sciences de l’Univers de Grenoble, and the Fundação para a Ciência e Tecnologia), to ESO and the Paranal staff, and to the many scientific and technical staff members in our institutions, who helped to make GRAVITY a reality. P.K. has received funding from the European Research Council (ERC) under the European Union's Horizon 2020 research and innovation program (projects CepBin, grant agreement 695099, and UniverScale, grant agreement 951549). A.A. and P.G. acknowledge the financial support provided by FCT/Portugal through grants PTDC/FIS-AST/7002/2020 and UIDB/00099/2020. F.W. has received funding from the European Union’s Horizon 2020 research and innovation programme under grant agreement No 10100471. This research has made use of the Jean-Marie Mariotti Center \texttt{Aspro} service. Based on observations collected at the European Southern Observatory under ESO programme(s): 
099.C-2017(A),
099.C-2017(B),
099.C-2017(C),
0100.C-0597(A),
0100.C-0597(B),
0100.C-0597(C),
0100.C-0608(A),
0100.C-0608(B),
0101.C-0121(A),
0101.C-0121(B),
0101.C-0121(C),
0102.C-0205(A),
0102.C-0205(B),
0102.C-0205(C),
0102.C-0211(A),
0102.C-0211(B),
0102.C-0211(C),
0103.C-0183(A),
0103.C-0183(B),
0103.C-0183(C),
0103.C-0192(A),
0103.C-0183(A),
0103.C-0183(B),
0103.C-0183(C),
0104.C-0161(A),
0104.C-0161(B),
0104.C-0161(C),
0104.C-0161(D),
105.20KT.001,
105.20KT.002,
105.20KT.003,
106.215T.001,
106.215T.002,
106.215T.003,
109.2310.001,
109.2310.002,
109.2310.003,
109.2310.004,
110.23TC.001,
110.23TC.002,
110.23TC.003,
111.24KS.001,
111.24KS.002,
111.24KS.003,
111.24KS.004,
111.24KS.005,
111.24KS.006,
112.25GN.001,
112.25GN.002,
112.25GN.003.
This research has made use of Astropy\footnote{Available at \url{http://www.astropy.org/}}, a community-developed core Python package for Astronomy \citepads{2013A&A...558A..33A,2018AJ....156..123A}, the Numpy library \citepads{Harris20}, the Scipy library \citepads{scipy} and the Matplotlib graphics environment \citepads{2007CSE.....9...90H}.
\end{acknowledgements}

%
%

\bibliographystyle{aa} 
\bibliography{aanda}      

\begin{appendix}

\section{Swap observations}
\label{annex1}

The goal of the swap sequence is to measure the phases of Eq.\ref{eq:astrom_phase} with two opposite signs for the field component, while keeping constant the zero-point of the metrology. This is done using the internal K-mirror of GRAVITY fiber coupler \citepads{2014SPIE.9146E..23P}, which allows us to rotate the field before the injection of light in the fibers, given the separation of the stars and the fact that both components are equally bright and can be equally used as a fringe-tracking star. 

The two phases which are measured for each step of the swap sequence are: 

\begin{equation}
\label{eq:swap}
\begin{cases}
\phi_{1} &= \frac{2\pi}{\lambda}(\overrightarrow{s}_B-\overrightarrow{s}_A)\cdot\overrightarrow{B}_{NAB,1}  + \phi_0(\lambda) \\
\phi_{2} &= \frac{2\pi}{\lambda}(\overrightarrow{s}_A-\overrightarrow{s}_B)\cdot\overrightarrow{B}_{NAB,2}  + \phi_0(\lambda)
\end{cases}
\end{equation}

For each baseline, the metrology zero point can finally be extracted from the combination of Eq.\ref{eq:swap} pointings:

\begin{equation}
\label{}
\begin{aligned}
\phi_0(\lambda) &= \frac{1}{2} \, \mathrm{arg} \left[ \; \tilde{V}_{FT,1} \, e^{-\frac{2\pi}{\lambda}(\overrightarrow{s}_B-\overrightarrow{s}_A) \overrightarrow{B}_{NAB,1}} \, \cdot \, \tilde{V}_{FT,2} \, e^{-\frac{2\pi}{\lambda}(\overrightarrow{s}_A-\overrightarrow{s}_B) \, \overrightarrow{B}_{NAB,2}} \; \right] \\
&= \frac{1}{2} \mathrm{arg} \left[ \, \tilde{V}_{FT,1}  \cdot \tilde{V}_{FT,2} \, \right] -  \frac{\pi}{\lambda} \, (\overrightarrow{s}_B -\overrightarrow{s}_A) \cdot \left( \overrightarrow{B}_{NAB,1} - \overrightarrow{B}_{NAB,2}\right)
\end{aligned}
\end{equation}
with $\tilde{V}_{FT,k}$ the complex visibility amplitude measured by the fringe-tracker. The metrology zero-point is then included in Eq.\ref{eq:swap}, which provides the relative separation $\overrightarrow{s}_B -\overrightarrow{s}_A$ of components A and B.

\section{Orbit fitting}
\label{annex2}
The stellar orbit and the planetary orbits are both defined using the Thiele-Innes elements, related to the four orbital orbital elements ($a$, $e$, $\omega$, $\Omega$). We followed the convention in \citetads{2023A&A...674A..34G, 2023A&A...674A...9H}:

\begin{equation}
\begin{cases}
    A = \cos\Omega\cos\omega - \sin\Omega\sin\omega\cos i \\
    B = \sin\Omega\cos\omega + \cos\Omega\sin\omega\cos i \\
    F = - \cos\Omega\sin\omega - \sin\Omega\cos\omega\cos i \\
    G = - \sin\Omega\sin\omega + \cos\Omega\cos\omega\cos i \\
\end{cases}
\end{equation}

The projected separation of the secondary with respect to the primary is then defined as:

\begin{equation}
\begin{cases}
    \Delta\delta = a\left(AX+FY\right) \\
    \Delta\alpha^* = a\left(BX+GY\right)
\end{cases}
\end{equation}

where X and Y are the elliptical rectangular coordinates:

\begin{equation}
\begin{cases}
    X = \cos E - e \\
    Y = \left(\sin E\right)\sqrt{1-e^2} 
\end{cases}
\end{equation}

with $E$ as the eccentric anomaly, obtained by solving the Kepler equation.

\begin{figure}
   \centering
   \includegraphics[width=1.1\hsize]{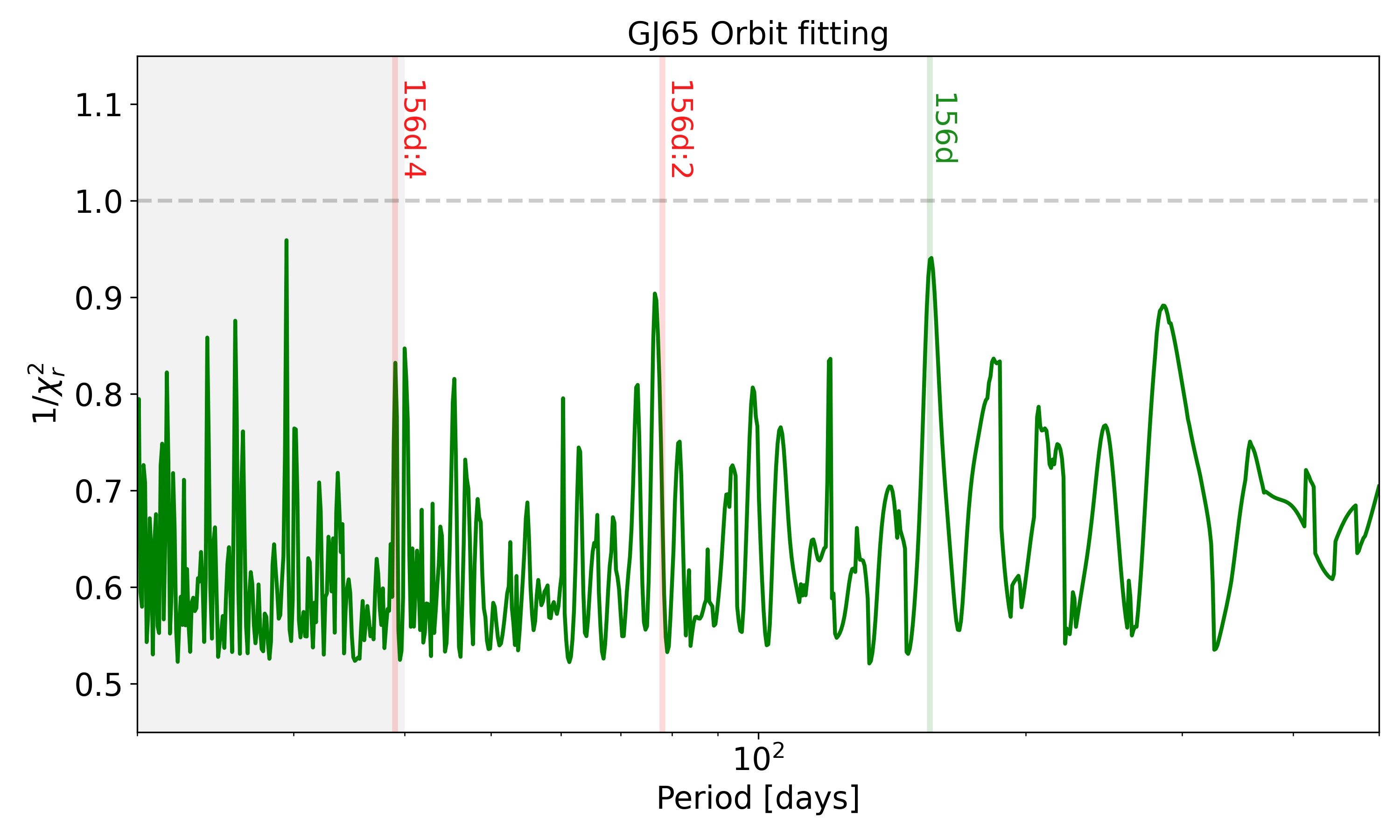}
      \caption{Period search of GJ65 residuals, with all orbital parameters free for each (fixed) period and including covariances, shown with inverse reduced chi-square in y-axis. The best period, $p=156\,\mathrm{d}$, coincides with the peak of the Lomb-Scargle periodogram. 
              }
         \label{FigPeriodSearch}
\end{figure}

\section{Stability domain}
\label{sec:stability_annex}
\begin{figure}[h]
   \centering
   \includegraphics[width=0.8\hsize]{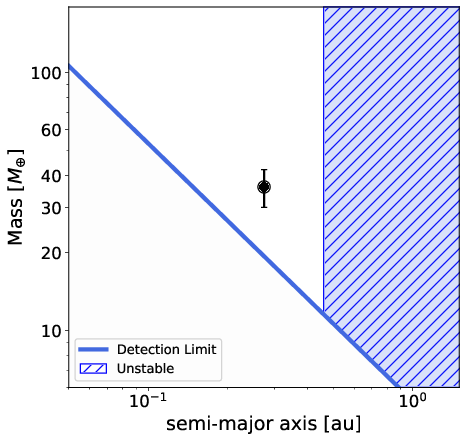}
      \caption{Parameter space probed by GRAVITY observations. The stability region is computed from \cite{1999AJ....117..621H}.}
         \label{FigPeriod}
   \end{figure}

We use the stability criterion of test particles derived in \citetads{1999AJ....117..621H} (Eq. 1), which we reproduce here:

\begin{equation}
    a_c = \left[
    0.464 - 0.380 \mu 
    -0.631 e + 0.586\mu e + 0.150e^2 - 0.198 \mu e^2
    \right] a_b
\end{equation}

with $a_c$, as the critical semi-major axis defining the outer edge of the stability region, $\mu=m_B/m_A$ the mass ratio of the binary, $e$ the eccentricity of the binary, and $a_b$ its semi-major axis. 
The Neptune-mass candidate as well as the stability region of the system are shown in Figure \ref{FigPeriod}.

\section{System architecture}

\begin{figure*}
   \centering
   \includegraphics[width=0.75\hsize]{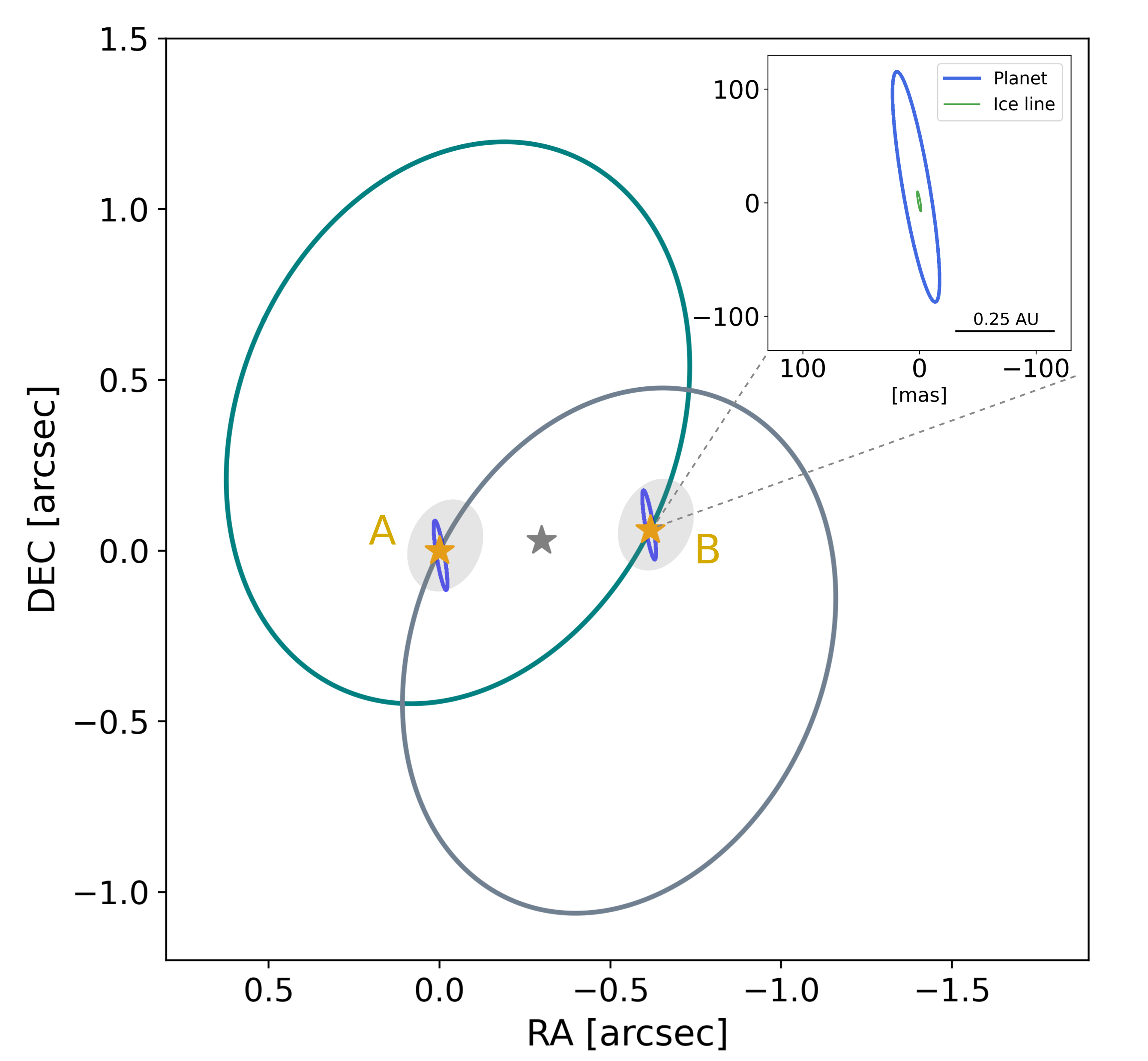}
    \caption{Planet candidate and barycentric ellipsis of the A and B stars in GJ65. The gray-shaded disk corresponds to the region of stability, shown in the orbital plane of the binary. The blue orbit corresponds to the planet in either star A and star B hypothesis. The inset shows the on-sky orbit around GJ65 B and the location of the ice-line (green) in the orbital plane of the planetary system.}
         \label{Fig_orbitAnnex_Planet}
\end{figure*}

\newpage

\begin{figure*}
   \centering
   \includegraphics[width=0.95\hsize]{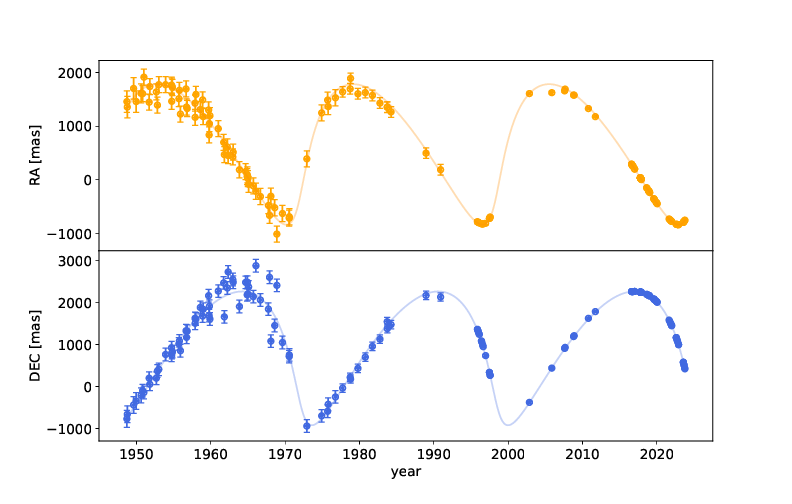}
    \caption{Relative angular separation of the stars A and B in GJ65.}
         \label{Fig_orbitAnnex_AM}
\end{figure*}

\begin{figure*}
   \centering
   \includegraphics[width=0.95\hsize]{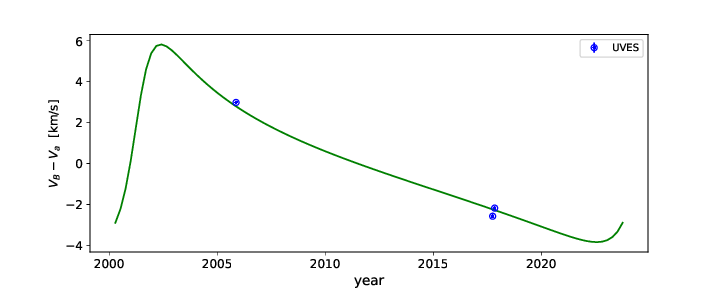}
    \caption{Relative radial velocity of stars A and B in GJ65.}
         \label{Fig_orbitAnnex_RV}
\end{figure*}

\end{appendix}

\end{document}